\documentclass{article}

\usepackage[cp1251]{inputenc}
\usepackage[english]{babel}
\usepackage{anysize}
\usepackage{amsmath}
\usepackage{amsfonts}
\usepackage{lscape}
\usepackage{subfigure}
\usepackage[pdftex]{graphicx}
\usepackage[usenames]{color}
\usepackage{colortbl}
\usepackage{amssymb}
\graphicspath{{images/}}
\numberwithin{equation}{section}
\long\gdef\comment#1{}

\begin{document}

\title{Interaction of orthogonal-polarized waves in 1D metamaterial with
Kerr nonlinearity}
\author{\vspace{6pt} Dmitrii Ampilogov, Sergey Leble \\
{\small Immanuel Kant Baltic Federal University, Al. Nevsky st 14,
Kaliningrad, Russia}}
\maketitle

\begin{center}
Abstract
\end{center}

 A theoretical study of wave propagation in 1D
metamaterial is presented. A system of nonlinear evolution equation for electromagnetic waves  with both polarizations account is derived by means of
projection operators method  for  general nonlinearity and
dispersion. The system describes interaction of opposite directed waves with a given polarization. The particular case of Kerr nonlinearity and Drude
dispersion is considered. In such approximation it results in the correspondent systems of nonlinear equations that
generalizes the Sch\"{a}fer-Wayne one. Particular solutions in case of
slow-varying envelopes are found, plotted and analyzed in gigahertz range.
Travelling wave solution for the system of equation of interaction of orthogonal-polarized waves is also obtained and the correspondent nonlinear dispersion relations are written in explicit form.

\section*{ Introduction}

\subsection*{ On nonlinear pulse propagation theory}

The nonlinear behavior of electromagnetic (EM) wave propagation depends on
relations between the field and induced polarization. It is obvious that it
is necessary to use either numerical scheme or approximations to obtain an
analytical solution of a nonlinear problem. The first successful approach of
such reduction was use the set of slowly varying envelopes. The simplest
model scalar equation for a directed wave propagation, based on this
approach, have a form of the nonlinear Schr\"{o}dinger equation, derived by
Zakharov in 1968 \cite{Zakh}. Its integrability \cite{ZS} made the model
very attractive because the rich "zoo" of the equation explicit solutions
\cite{DL}.

A natural step of integrable generalization of such model lies in a plane of
better approximations of dispersion, dissipation \cite{P} and nonlinearity
(modified nonlinear Schr\"{o}dinger (MNS) equation, see e.g. \cite{DL}),
that allows to extend pulse durations down to picoseconds.

There are plenty of alternative ideas on the few-cycle pulse soliton-type
description in different media \cite{SU}

The next step of this movement to the ultrashort pulses description that
maintains integrability is made in works of Sh\"{a}fer-Wayne \cite{SW,SW2}.
The Short Pulse Equation (SPE) again relates to unidirectional propagation
for which a special kind of dispersion law and nonlinearity action was
account in a rescaled evolution. In 2017 Z. Zhaqilao et al. derived an
N-fold Darboux transformation from the Lax pair of the two-component short
pulse system with loop. \cite{Zhaqilao}

A generalization that allows to include description and interaction of
opposite directed waves is connected with idea of joint account of the
correspondent spaces of "hybrid" electric-magnetic amplitudes \cite{K,L}.
The projecting operators (PO) method \cite{L} works at arbitrary dispersion
and nonlinearity. Similar universality demonstrates a method of \cite{Dobr}.
The PO technique gives systematic transition to hybrid fields with
simultaneous superposition of nonlinear terms, that effectively approximate
weak nonlinearity, arriving at the mentioned celebrated model equations at
the subspaces of directed waves \cite{KL}. The field hybridization may
account ab initio dispersion and dissipation \cite{P} and nonlinearity by
iteration procedure \cite{Pe}.

Next, natural for electromagnetic field step accounts polarization and leads
to double component vector equations \cite{KuLe}, similar vector equation is
studied in \cite{PKB}. Both direction and polarization studied theoretically
and, of importance, experimentally in \cite{Pi}.

\subsection*{ On Drude model}
One of important applications of such model relates to metamaterials that
characterized by negative values of the parameters $\varepsilon$
and $\mu$, that must be anomalous dispersive, i.e., their permittivity and
permeability must be frequency dependent, otherwise they would not be causal
\cite{Ziol}. The two-time derivative Lorentz material model encompasses the
metamaterial models most commonly discussed; it has the frequency domain
susceptibility \cite{Ziol2}:
\begin{equation}
\chi=\frac{\omega_{p}\chi_{a}+i\omega_{p}\chi_{\beta}\omega-\chi_{\gamma
}\omega}{\omega_{0}+i\omega\Gamma-\omega},
\end{equation} 
its particular case is the 2TDLM model, which produces a resonant response at $\omega=\omega_{0}$ when $\Gamma=0$. It recovers the Drude model when this resonant frequency goes to zero, and the constants $ \chi_{\alpha}=1$, $\chi_{\beta}=\chi_{\gamma}=0$. For this model Kanattsikov and Pietrzyk pointed, that the propagation of ultra-short pulses could be described by short pulse Sch\"{a}fer - Wayne equation \cite{Kan2}

\subsection*{ Aim and scope}

It's continued systematic application of projecting approach, originated
from \cite{L} for 1D metamaterial with account of both polarizations of the
EM wave. The technique and results of our previous work \cite{AmLe} on
nonlinear evolution equations of opposite directed waves with one
polarization in Drude 1D-metamaterial has been developed.

In this paper application projecting operators method for this case is
demonstrated. Obtained nonlinear equation for metamaterial with our previous
results and vector SPE is compared. On base of resulting equations, the wave
packets for linear and nonlinear cases are studied. Preliminary results got in \cite{AmLETASK}.

For ordinary nonlinear Kerr material Kaplan in 1983 showed saving the
arrangements of polarizations \cite{Kaplan}. But for metamaterial the
situation is different, as shown in our work \cite{AmLe} for unique
polarization. We discover the change of arrangements of wave modes. Now, the
questions to be answered: what happens with account of polarization? And how
looks interactions of all four modes in metamaterial? The content of the
paper is following

\begin{itemize}
\item Sec.1: Statement the boundary regime problem,

\item Sec.2: The 4x4 matrix projection operators with arbitrary dispersion
account are built for the case of two polarizations,

\item Sec.3 : Derivation the general linear system of equations for two left
and two right waves with orthogonal polarizations.

\item Sec.4 :The general nonlinear system of equations for the left and
right waves and two polarizations is obtained.

\item Sec. 5: For the particular case of the Kerr nonlinearity within
approximate Drude dispersion got the novel system of short pulse equations
that is reduced to the Sh\"{a}fer-Wayne one for unique polarization.

\item In the Sec 6 attention have been fixed to wave trains, starting from
linear and, taking into account nonlinearity, obtain a plane wave with
amplitude dependent wavelength.
\end{itemize}

\section{ Maxwell's equations. Boundary regime problem}

The starting point is the Maxwell equations for linear isotropic dispersive
dielectric media, in the SI unit system:
\begin{eqnarray}
\text{div}\vec{D}(\vec{r},t) &=&0,  \label{eq:Maxwell_D} \\
\text{div}\vec{B}(\vec{r},t) &=&0,  \label{eq:Maxwell_B} \\
\text{rot}\vec{E}(\vec{r},t) &=&-\frac{\partial \vec{B}(\vec{r},t)}{\partial
t},  \label{eq:Faradey} \\
\text{rot}\vec{H}(\vec{r},t) &=&\frac{\partial \vec{D}(\vec{r},t)}{\partial
t }.  \label{eq:Amper}
\end{eqnarray}
Restricting ourself to a one-dimensional model, similarly to Sh\"{a}fer,
Wayne \cite{SW}, where the $x$-axis is chosen as the direction of a wave
propagation assuming zero longitudinal field components allows to write the
Maxwell equations with arbitrary polarization account :
\begin{equation}
\begin{array}{cc}
\frac{\partial D_{y}}{\partial t} & =-\frac{\partial H_{z}}{\partial x}, \\
\frac{\partial D_{z}}{\partial t} & =\frac{\partial H_{y}}{\partial x}, \\
\frac{\partial B_{y}}{\partial t} & =\frac{\partial E_{z}}{\partial x}, \\
\frac{\partial B_{z}}{\partial t} & =-\frac{\partial E_{y}}{\partial x},%
\end{array}
\label{eq:1DMax2}
\end{equation}
To close the system \eqref{eq:1DMax2} we need to add material
relations: where $\widehat{\mu }$ and $\widehat{\varepsilon }$ are integral
convolution-type operators \cite{AmLe}:
\begin{eqnarray}
\widehat{\varepsilon }\psi (x,t) &=&\int\limits_{-\infty }^{\infty }\tilde{
\varepsilon}(t-s)\psi (x,s)ds,  \label{opEps} \\
\widehat{\mu }\psi (x,t) &=&\int\limits_{-\infty }^{\infty }\tilde{\mu}
(t-s)\psi (x,s)ds,
\end{eqnarray}
with kernels
\begin{equation}
\tilde{\varepsilon}(t-s) =\frac{\varepsilon _{0}}{2\pi }
\int\limits_{-\infty }^{\infty }\varepsilon (\omega )\exp (i\omega
(t-s))d\omega , \tilde{\mu}(t-s) =\frac{\mu _{0}}{2\pi }\int\limits_{-\infty }^{\infty
}\mu (\omega )\exp (i\omega (t-s))d\omega .
\end{equation}
Hence operator form of the equation \eqref{eq:1DMax2} is:
\begin{eqnarray}
\partial _{t}(\widehat{\varepsilon }E_{y,z}) &=&\mp \partial _{x}(\widehat{
\mu }^{-1}B_{z,y}),  \label{eq:1DMax3} \\
\nonumber \partial _{t}B_{y,z} &=&\pm \partial _{x}E_{z,y}.
\end{eqnarray}
Here it's marked:
\begin{equation}
\frac{\partial }{\partial x} \equiv \partial _{x}, \frac{\partial }{\partial t} \equiv \partial _{t}.
\end{equation}
Adding boundary conditions to state the problem:
\begin{equation}
E_{y,z}(0,t)=j_{y,z}(t),\,\,\,B_{y,z}(0,t)=\ell _{y,z}(t),  \label{bc}
\end{equation}
$j_{i}$ and $\ell _{i}$ are arbitrary functions, continued to the half space
$t<0$ antisymmetricaly:
\begin{equation}
j_{i}(-t)=-j_{i}(t),\ell _{i}(-t)=-\ell _{i}(t),i=y,z.
\end{equation}

\section{ Dynamic projecting operators}

Doing the Fourier transformations like in \cite{AmLe} and plugging them into
the system of equations \eqref{eq:1DMax3} one have the closed system:
\begin{equation}
\partial _{t}\int\limits_{-\infty }^{\infty }\varepsilon (\omega )\mathcal{E}
_{y,z}(x,\omega )\exp (i\omega t)d\omega =\mp \frac{1}{\mu _{0}\varepsilon
_{0}}\partial _{x}\int\limits_{-\infty }^{\infty }\frac{\mathcal{B}
_{z,y}(x,\omega )}{\mu (\omega )}\exp (i\omega t)d\omega .  \label{Max2BE}
\end{equation}
The inverse Fourier transformation yields in the four equations of %
\eqref{eq:1DMax3}, written in short form:
\begin{eqnarray}
\partial _{x}\mathcal{B}_{z,y} &=&\mp i\omega \mu _{0}\varepsilon _{0}\mu
(\omega )\varepsilon (\omega )\mathcal{E}_{y,z}.  \label{eq:ByEy} \\
\partial _{x}\mathcal{E}_{z,y} &=&\pm i\omega \mathcal{B}_{y,z}.
\label{eq:ByEy2}
\end{eqnarray} 
Let us define the column of the field components transform and matrix operator with obvious elements from (\ref{eq:ByEy}, \ref{eq:ByEy2})
\begin{equation}
\tilde{\Psi}=\left(
\begin{array}{c}
\mathcal{B}_{z} \\
\mathcal{B}_{y} \\
\mathcal{E}_{z} \\
\mathcal{E}_{y}%
\end{array}
\right) ,
\mathcal{L}=\left(
\begin{array}{cc}
\widehat{0} & \mathcal{L}_{1} \\
\mathcal{L}_{2} & \widehat{0}%
\end{array}
\right) , \label{koshi-kt}
\end{equation}
arriving at
\begin{equation}
\partial _{x}\tilde{\Psi}=\mathcal{L}\tilde{\Psi}.  \label{mxeqw}
\end{equation}
Doing projection operators technique, similarly with \cite{AmLe}, finally we obtain four matrix projecting operators in $t-$representation by the standard general formula:
\begin{equation}
\mathbf{P}^{(1)}=\frac{1}{2}\left(
\begin{array}{cccc}
1 & 0 & 0 & -\widehat{a} \\
0 & 0 & 0 & 0 \\
0 & 0 & 0 & 0 \\
-\widehat{a}^{-1} & 0 & 0 & 1%
\end{array}%
\right) ,
\mathbf{P}^{(2)}=\frac{1}{2}\left(
\begin{array}{cccc}
0 & 0 & 0 & 0 \\
0 & 1 & \widehat{a} & 0 \\
0 & \widehat{a}^{-1} & 1 & 0 \\
0 & 0 & 0 & 0%
\end{array}%
\right) , \label{projectors}
\end{equation}%
\begin{equation}
\mathbf{P}^{(3)}=\frac{1}{2}\left(
\begin{array}{cccc}
1 & 0 & 0 & \widehat{a} \\
0 & 0 & 0 & 0 \\
0 & 0 & 0 & 0 \\
\widehat{a}^{-1} & 0 & 0 & 1%
\end{array}%
\right) ,
\mathbf{P}^{(4)}=\frac{1}{2}\left(
\begin{array}{cccc}
0 & 0 & 0 & 0 \\
0 & 1 & -\widehat{a} & 0 \\
0 & -\widehat{a}^{-1} & 1 & 0 \\
0 & 0 & 0 & 0%
\end{array}%
\right).
\end{equation}%
Projectors $\mathbf{P}^{(1,2)}$ correspond to $\lambda _{1}$ and two other
ones - to $\lambda _{2}$.\newline
Operators $\widehat{a}$, $\widehat{a}^{-1}$ are defined as \cite{AmLe}:
\begin{eqnarray}
\widehat{a}\eta (x,t) &=&\frac{1}{2\pi }\int\limits_{-\infty }^{\infty }%
\left[ \eta (x,\tau )\int\limits_{-\infty }^{\infty }a(\omega )\exp (i\omega
(t-\tau ))d\omega \right] d\tau ,  \label{hata} \\
\widehat{a}^{-1}\xi (x,t) &=&\frac{1}{2\pi }\int\limits_{-\infty }^{\infty }%
\left[ \xi (x,\tau )\int\limits_{-\infty }^{\infty }\frac{1}{a(\omega )}\exp
(i\omega (t-\tau ))d\omega \right] d\tau ,  \notag
\end{eqnarray}%
where $a(\omega )$ is positive solution of the quadratic equation \eqref{qv}:
\begin{equation}
\mu _{0}\varepsilon _{0}\varepsilon (\omega )\mu (\omega )\equiv
c^{-2}\varepsilon (\omega )\mu (\omega )\equiv a^{2}(\omega ),
\label{def:a(w)}
\end{equation}%
where $c=\frac{1}{\sqrt{\varepsilon _0\mu_0}}$ is the velocity of light in vacuum.

\section{ Separated equations and definition for left and right waves}

Let's return to the time-domain. Let's write the matrix equation %
\eqref{mxeqw} in this representation:
\begin{equation}
\partial _{x}\Psi =\widehat{L}\Psi ,  \label{relPsi}
\end{equation}%
where
\begin{equation}
\Psi =\left(
\begin{array}{c}
B_{z} \\
B_{y} \\
E_{z} \\
E_{y}%
\end{array}%
\right) ,  \label{psit}
\end{equation}%
\begin{equation}
\widehat{L}=\left(
\begin{array}{cccc}
0 & 0 & 0 & -\partial _{t}\widehat{a}^{2} \\
0 & 0 & \partial _{t}\widehat{a}^{2} & 0 \\
0 & \partial _{t} & 0 & 0 \\
-\partial _{t} & 0 & 0 & 0%
\end{array}%
\right) .  \label{L-need}
\end{equation}%
Action of projectors $P^{(1)}$ and $P^{(2)}$ on \eqref{relPsi} yields to
hybrid waves $\Pi _{1}$ and $\Pi _{2}$:
\begin{equation}
\Pi _{1}=\frac{1}{2}(B_{z}-\widehat{a}E_{y}),  \label{def:leftwave1}
\end{equation}%
\begin{equation}
\Pi _{2}=\frac{1}{2}(B_{y}+\widehat{a}E_{z}).  \label{def:rightwave2}
\end{equation}%
Action of projectors $P^{(3)}$ and $P^{(4)}$ on \eqref{relPsi} yields to
hybrid waves $\Lambda _{1}$ and $\Lambda _{2}$:
\begin{equation}
\Lambda _{1}=\frac{1}{2}(B_{z}+\widehat{a}E_{y}),  \label{def:rightwave1}
\end{equation}%
\begin{equation}
\Lambda _{2}=\frac{1}{2}(B_{y}-\widehat{a}E_{z}).  \label{def:leftwave2}
\end{equation}%
Waves $\Pi _{1}$ and $\Lambda _{1}$ introduced for the case of unique
polarization \cite{AmLe} and describe a propagation of $y-$polarization. The
other two do the same for $z-$polarization.
\begin{eqnarray}
\partial _{x}\Pi _{1} =\partial _{t}\widehat{a}\Pi _{1},
\partial _{x}\Pi _{2} =\partial _{t}\widehat{a}\Pi _{2}, \\
\partial _{x}\Lambda _{1} &=& -\partial _{t}\widehat{a}\Lambda _{1}, \\
\partial _{x}\Lambda _{2} &=&-\partial _{t}\widehat{a}\Lambda _{2}.
\end{eqnarray}%
Using definitions of (\ref{def:leftwave1}- \ref{def:rightwave2} ) and %
\eqref{bc} we derive boundary regime conditions for left and right waves:
\begin{eqnarray}
\Lambda _{1}(0,t) &=&\frac{1}{2}(B_{z}(0,t)+\widehat{a}E_{y}(0,t))=\frac{1}{2%
}(k_{z}(t)+\widehat{a}j_{y}(t)), \\
\Lambda _{2}(0,t) &=&\frac{1}{2}(B_{y}(0,t)-\widehat{a}E_{z}(0,t))=\frac{1}{2%
}(k_{y}(t)-\widehat{a}j_{z}(t)), \label{bcq} \\
\Pi _{1}(0,t) &=&\frac{1}{2}(B_{z}(0,t)-\widehat{a}E_{y}(0,t))=\frac{1}{2}%
(\ell _{z}(t)+\widehat{a}j_{y}(t)), \\ \Pi _{2}(0,t) &=&\frac{1}{2}(B_{y}(0,t)+\widehat{a}E_{z}(0,t))=\frac{1}{2}%
(\ell _{y}(t)+\widehat{a}j_{z}(t)).
\end{eqnarray}

\section{General nonlinearity account}

Let us consider a nonlinear problem. The starting point is the Maxwell's
equations \eqref{eq:1DMax2} again with generalized nonlinear material
relations:
\begin{eqnarray}
D_{i} &=&\widehat{\varepsilon }E_{i}+P_{i}^{(NL)}, \\
B_{i} &=&\widehat{\mu }H_{i}+M_{i}^{(NL)},i=y,z.  \notag  \label{nlmat}
\end{eqnarray}
$P_{NL}$ - nonlinear part of polarization ($M_{NL}$ - one for
magnetization). Linear parts of polarization and magnetization have already
been taken into account. In time-domain, a closed nonlinear version of %
\eqref{eq:1DMax3} is:
\begin{eqnarray}
\partial _{t}(\widehat{\varepsilon }E_{y,z})+\partial _{t}P_{y,z}^{(NL)}
&=&\mp \partial _{x}\widehat{\mu }^{-1}B_{z,y}\mp \partial _{x}\widehat{\mu }%
^{-1}M_{z,y}^{(NL)},  \label{eq:1DMax4} \\
\partial _{t}B_{z,y} &=&-\partial _{x}E_{y,z}.
\end{eqnarray}%
Action of operator $\widehat{\mu }$ on the first pair of equations of system %
\eqref{eq:1DMax4} and use of the same notations $\Psi $ and $\widehat{L}$
from (\ref{psit}, \ref{L-need}) once more, produce a nonlinear analogue of
the matrix equation \eqref{relPsi}:
\begin{equation}
\partial _{x}\Psi -\widehat{L}\Psi =\partial _{x}\left(
\begin{array}{c}
M_{z}^{(NL)} \\
M_{y}^{(NL)} \\
0 \\
0%
\end{array}%
\right) -\widehat{\mu }\partial _{t}\left(
\begin{array}{c}
P_{y}^{(NL)} \\
-P_{z}^{(NL)} \\
0 \\
0%
\end{array}%
\right) .  \label{relPsinl}
\end{equation}%
In the r.h.s there is a vector of nonlinearity for case of the opposite
directed 1D-waves:
\begin{equation}
\mathbb{N}=\left(
\begin{array}{c}
\partial _{x}M_{z}^{(NL)}-\widehat{\mu }\partial _{t}P_{y}^{(NL)} \\
\partial _{x}M_{y}^{(NL)}+\widehat{\mu }\partial _{t}P_{z}^{(NL)} \\
0 \\
0%
\end{array}%
\right) .  \label{eq:vectorN}
\end{equation}

Next, acting by operators $\widehat{\mathbf{P}}^{(1,2,3,4)}$ %
\eqref{projectors} on the Eq. \eqref{relPsinl} one can find:
\begin{eqnarray}
\partial _{x}\Pi _{1}-\partial _{t}\widehat{a}\Pi _{1} &=&-\frac{1}{2}%
(\partial _{x}M_{z}^{(NL)}-\widehat{\mu }\partial _{t}P_{y}^{(NL)}),
\label{eqs:systemofwavesnonlinear} \\
\partial _{x}\Pi _{2}-\partial _{t}\widehat{a}\Pi _{2} &=&\frac{1}{2}%
(\partial _{x}M_{y}^{(NL)}+\widehat{\mu }\partial _{t}P_{z}^{(NL)}), \\
\partial _{x}\Lambda _{1}+\partial _{t}\widehat{a}\Lambda _{1} &=&\frac{1}{2}%
\left( \partial _{x}M_{z}^{(NL)}-\widehat{\mu }\partial
_{t}P_{y}^{(NL)}\right) , \\
\partial _{x}\Lambda _{2}+\partial _{t}\widehat{a}\Lambda _{2} &=&-\frac{1}{2%
}\left( \partial _{x}M_{y}^{(NL)}+\widehat{\mu }\partial
_{t}P_{z}^{(NL)}\right) .
\end{eqnarray}%
Generally the r.h.s. of each equation \eqref{eqs:systemofwavesnonlinear}
depends on the field vectors $\vec{E},\vec{B}$, that should be presented in
terms of the fields $\vec{\Pi},\vec{\Lambda}$ to close the system. The
vectors components are expressed by means of the inverse transformation of ( %
\ref{def:leftwave1} - \ref{def:rightwave2}).

\section{ Kerr nonlinearity account for lossless Drude metamaterials}

\subsection{ Equations of interaction of the waves via Kerr effect}

For nonlinear Kerr materials \cite{Christos2}, the third-order nonlinear
part of polarization \cite{KuLe,Christos2} has the form:
\begin{equation*}
P_{y,z}^{(NL)}=\varepsilon _{0}\chi^{(3)}(E_{y,z}^{3}+E_{y,z}E_{z,y}^{2}).
\end{equation*}
From \eqref{eq:vectorN}, deleting magnetic nonlinearity, one can find the
vector $N$:

\begin{equation}
\mathbb{N}=-\varepsilon _{0}\widehat{\mu }\chi _{e}^{(3)}\partial _{t}\left(
\begin{array}{c}
E_{y}^{3}+E_{y}E_{z}^{2} \\
-E_{z}^{3}-E_{z}E_{y}^{2} \\
0 \\
0%
\end{array}%
\right) .  \label{eq:vectorNKerr}
\end{equation}%
Account for the definitions of the hybrid fields as \eqref{def:leftwave1}
gives:
\begin{eqnarray}
E_{y} &=&-\frac{1}{2}\widehat{a}^{-1}(\Pi _{1}-\Lambda _{1}), \\
E_{z} &=&-\frac{1}{2}\widehat{a}^{-1}(\Pi _{2}-\Lambda _{2}).
\end{eqnarray}

\bigskip

The system for left and right waves with two polarizations equations in a
medium with Kerr nonlinearity:

\begin{eqnarray}
\partial _{x}\Pi _{1}-\partial _{t}\widehat{a}\Pi _{1} &=&-\frac{1}{16}%
\varepsilon _{0}\chi _{e}^{(3)}\widehat{\mu }\partial _{t}[(\widehat{a}%
^{-1}(\Pi _{1}-\Lambda _{1}))^{3}+\widehat{a}^{-1}(\Pi _{1}-\Lambda _{1})(%
\widehat{a}^{-1}(\Pi _{2}-\Lambda _{2}))^{2}],  \label{eqs:nlsystemofwave} \\
\partial _{x}\Pi _{2}-\partial _{t}\widehat{a}\Pi _{2} &=&-\frac{1}{16}%
\varepsilon _{0}\chi _{e}^{(3)}\widehat{\mu }\partial _{t}[(\widehat{a}%
^{-1}(\Pi _{2}-\Lambda _{2}))^{3}+\widehat{a}^{-1}(\Pi _{2}-\Lambda _{2})(%
\widehat{a}^{-1}(\Pi _{1}-\Lambda _{1}))^{2}] \\
\partial _{x}\Lambda _{1}+\partial _{t}\widehat{a}\Lambda _{1} &=&\frac{1}{16%
}\varepsilon _{0}\chi _{e}^{(3)}\widehat{\mu }\partial _{t}[(\widehat{a}%
^{-1}(\Pi _{1}-\Lambda _{1}))^{3}+\widehat{a}^{-1}(\Pi _{1}-\Lambda _{1})(%
\widehat{a}^{-1}(\Pi _{2}-\Lambda _{2}))^{2}], \\
\partial _{x}\Lambda _{2}+\partial _{t}\widehat{a}\Lambda _{2} &=&\frac{1}{16%
}\varepsilon _{0}\chi _{e}^{(3)}\widehat{\mu }\partial _{t}[(\widehat{a}%
^{-1}(\Pi _{2}-\Lambda _{2}))^{3}+\widehat{a}^{-1}(\Pi _{2}-\Lambda _{2})(%
\widehat{a}^{-1}(\Pi _{1}-\Lambda _{1}))^{2}].
\end{eqnarray}%
In unidirectional case with $\Pi _{1},\Pi _{2}=0$ one can obtain the system,
that describe interaction between hybrid fields with different
polarizations:
\begin{eqnarray}
\partial _{x}\Lambda _{1}+\partial _{t}\widehat{a}\Lambda _{1} &=&-\frac{1}{%
16}\varepsilon _{0}\widehat{\mu }\chi _{e}^{(3)}\partial _{t}[(\widehat{a}%
^{-1}\Lambda _{1})^{3}+\widehat{a}^{-1}\Lambda _{1}(\widehat{a}^{-1}\Lambda
_{2})^{2}],  \label{unidirectionalsystemofwaves} \\
\partial _{x}\Lambda _{2}+\partial _{t}\widehat{a}\Lambda _{2} &=&-\frac{1}{%
16}\varepsilon _{0}\widehat{\mu }\chi _{e}^{(3)}\partial _{t}[(\widehat{a}%
^{-1}\Lambda _{2})^{3}+\widehat{a}^{-1}\Lambda _{2}(\widehat{a}^{-1}\Lambda
_{1})^{2}].
\end{eqnarray}%
Because of propagation in one direction, it's useful to mark $\Lambda _{1}$
and $\Lambda _{2}$ as: Applying Drude model, approximately write
\begin{equation}
\widehat{a}^{-1}\eta (x,t)\approx \frac{c}{pq}\partial _{t}^{2}\eta (x,t),
\label{expansiona-1}
\end{equation}%
\begin{equation}
\widehat{\mu }\eta (x,t)\approx -\mu _{0}q^{2}\partial _{t}^{-2}\eta (x,t),
\label{expansiona-1}
\end{equation}%
(see again \cite{AmLe} for details), plugging $\varepsilon _{0}\mu
_{0}=c^{-2}$,
\begin{eqnarray}
\frac{c}{pq}\partial _{x}\Lambda _{1}+\partial _{t}^{-1}\Lambda _{1} &=&%
\frac{1}{16}\chi _{e}^{(3)}\frac{c^{2}}{p^{4}q^{2}}\partial
_{t}^{-1}[(\partial _{t}^{2}\Lambda _{1})^{3}+\partial _{t}^{2}\Lambda
_{1}(\partial _{t}^{2}\Lambda _{2})^{2}], \\
\frac{c}{pq}\partial _{x}\Lambda _{2}+\partial _{t}^{-1}\Lambda _{1} &=&%
\frac{1}{16}\chi _{e}^{(3)}\frac{c}{p^{3}q}\partial _{t}^{-1}[(\partial
_{t}^{2}\Lambda _{2})^{3}+\partial _{t}^{2}\Lambda _{2}(\partial
_{t}^{2}\Lambda _{1})^{2}].  \label{unidirectionalsystemofwaves2}
\end{eqnarray}%
Differentiation on $t$ three times leads to equation:
\begin{eqnarray}
\frac{c}{pq}\partial _{xttt}\Lambda _{1}+\partial _{t}^{2}\Lambda _{1}
&=&\gamma ^{2}\partial _{t}^{2}[(\partial _{t}^{2}\Lambda _{1})^{3}+\partial
_{t}^{2}\Lambda _{1}(\partial _{t}^{2}\Lambda _{2})^{2}], \\
\frac{c}{pq}\partial _{xttt}\Lambda _{2}+\partial _{t}^{2}\Lambda _{2}
&=&\gamma ^{2}\partial _{t}^{2}[(\partial _{t}^{2}\Lambda _{2})^{3}+\partial
_{t}^{2}\Lambda _{2}(\partial _{t}^{2}\Lambda _{1})^{2}],
\label{unidirectionalsystemofwavesdff}
\end{eqnarray}%
where
\begin{equation}
\gamma ^{2}=\frac{1}{16}\chi _{e}^{(3)}\frac{c^{2}}{p^{4}q^{2}}.
\end{equation}%
Introducing new field functions $l_{1}$ and $l_{2}$ and variable $\chi $ as:
\begin{eqnarray}
l_{1} &\equiv &\gamma ^{-1}\partial _{t}^{2}\Lambda _{1}, \\ l_{2} &\equiv &\gamma ^{-1}\partial _{t}^{2}\Lambda _{2}, \\
\partial _{x} &=&\frac{pq}{c}\partial _{\chi },
\end{eqnarray}%
one can obtain the generalized SPE system :
\begin{eqnarray}
\partial _{\chi t}l_{1}+l_{1} &=&\partial _{t}^{2}(l_{1}^{3}+l_{2}^{2}l_{1}),
\label{pilambda} \\
\partial _{\chi t}l_{2}+l_{2} &=&\partial _{t}^{2}(l_{2}^{3}+l_{2}l_{1}^{2}).
\end{eqnarray}%
That's a generalization of Sch\"{a}fer-Wayne equation for case of
interaction of two left waves, which is one of objectives of this work.

\section{ Wave trains}

\subsection{ Linear wave packets for the right waves}

We consider the system \eqref{unidirectionalsystemofwaves2}, differentiated
on $t$:
\begin{eqnarray}
\frac{c}{pq}\partial _{xt}\Lambda _{1}+\Lambda _{1} &=&\frac{1}{16}\chi
_{e}^{(3)}\frac{c^{2}}{p^{4}q^{2}}[(\partial _{t}^{2}\Lambda
_{1})^{3}+\partial _{t}^{2}\Lambda _{1}(\partial _{t}^{2}\Lambda _{2})^{2}],
\\
\frac{c}{pq}\partial _{xt}\Lambda _{2}+\Lambda _{2} &=&\frac{1}{16}\chi
_{e}^{(3)}\frac{c^{2}}{p^{4}q^{2}}[(\partial _{t}^{2}\Lambda
_{2})^{3}+\partial _{t}^{2}\Lambda _{2}(\partial _{t}^{2}\Lambda _{1})^{2}].
\label{systemdifft}
\end{eqnarray}%
In linear case the r.h.s. will be equal 0. These equations are identical,
hence, take one of them :
\begin{equation}
\frac{c}{pq}\partial _{xt}\Lambda _{1}+\Lambda _{1}=0,  \label{linunid}
\end{equation}%
plugging the wavetrain solution, that we prepare for a comparison with
nonlinear case:
\begin{equation}
\Lambda _{1}=A(x,t)\exp [i(kx-\omega t)]+c.c..  \label{PiL}
\end{equation}%
Differentiating
\begin{equation}
\partial _{xt}\Lambda _{1}=A_{xt}\exp [i(kx-\omega t)]+ikA_{t}\exp
[i(kx-\omega t)]-i\omega A_{x}\exp [i(kx-\omega t)]-i\omega (ik)A\exp
[i(kx-\omega t)]+c.c.,  \label{dxt}
\end{equation}%
Putting the result in the equation \eqref{linunid}, assuming slow varying
amplitude:
\begin{equation}
A_{x}<<kA,A_{t}<<\omega A,  \label{approximationsA}
\end{equation}%
to kill the zeroth order term gives the dispersion relation:
\begin{equation}
k(\omega )=-\frac{pq}{c\omega },  \label{disp}
\end{equation}%
then, in the first order the equation arrives at
\begin{equation}
A_{t}-\frac{\omega }{k}A_{x}=0,  \label{lineareqenvshort}
\end{equation}%
Next, denoting
\begin{equation}
v_{g}=\frac{\omega }{k},  \label{vg}
\end{equation}%
after conventional change the variables:
\begin{equation}
\eta =t-\frac{x}{v_{g}},\xi =t+\frac{x}{v_{g}},  \label{xxi}
\end{equation}%
\begin{equation*}
\partial _{t}=\partial _{\xi }+\partial _{\eta },\partial _{x}=\frac{1}{v_{g}%
}\partial _{\xi }-\frac{1}{v_{g}}\partial _{\eta },A(x,t)\rightarrow \mathbb{%
A}(\eta ,\xi ),
\end{equation*}%
the \eqref{lineareqenvshort} trivializes as
\begin{equation}
2\mathbb{A}_{\eta }=0.
\end{equation}%
It's shown, the amplitude function $A$ is independent on $\eta $
\begin{equation*}
\mathbb{A}=f(\xi );A(x,t)=f\left( t+\frac{x}{v_{g}}\right) .
\end{equation*}%
Substituting this relation into \eqref{lineareqenvshort} leads to the
definition of $v_{g}$:
\begin{equation}
v_{g}=-\frac{\omega ^{2}c}{pq}  ,\label{def:vg}
\end{equation}
Sign "minus" means right direction of wave propagation. \newline
To fix the unique solution, it's necessary to add a boundary condition:
\begin{equation}
A(0,t)=A_{0}\exp\left[ -\left( \frac{t}{\tau}\right) ^{2}\right] ,
\label{bcA0}
\end{equation}
$\tau$ characterizes a width of wave packet and $\omega$ characterizes the
period of oscillation. Accounting the boundary regime \eqref{bcA0}, for the $%
\Lambda_1-$wave, propagated to the right the explicit formula is obtained:
\begin{equation}
\Lambda_{1}(x,t)=A_{0}\exp\left\{ -\left( \frac{t+\frac{x}{v_{g}}}{\tau}%
\right) ^{2}+i(kx-\omega t)\right\} +c.c.  \label{form}
\end{equation}
\comment{
\begin{figure}[h]
\begin{center}
\begin{minipage}[H]{0.4\linewidth}
\includegraphics[scale=0.4]{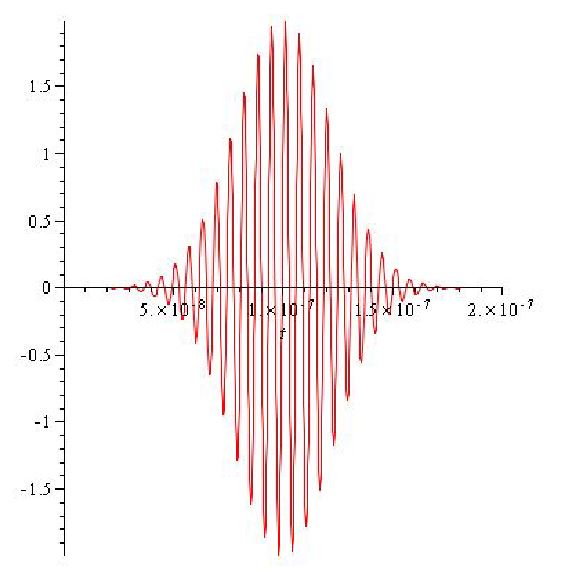}
\end{minipage}
\end{center}
\caption{Wave packet of $\Lambda_{1}$ for case $\protect\omega=10^{9}$ Hz, $%
x=0$ m, $v_{g}=10^{8}$ m/s, $\protect\tau=50T$ s}
\end{figure}}
The wavetrains with other polarization differs only by electric and magnetic
fields components numbers as it is prescribed by \eqref{def:leftwave2}. The
opposite directed waves are defined by \eqref{def:leftwave1}, %
\eqref{def:rightwave2}, its formulas differ from \eqref{form} only by signs
by $\frac{x}{v_{g}}$.

\subsection{ Dispersionless nonlinear equations for envelopes}

We consider the system \eqref{systemdifft}. For $\Lambda_{1}$ and $%
\Lambda_{2}$ in the wavetrain form with the frequency chosen by boundary
condition as in linear case:
\begin{eqnarray}
\Lambda_{1} & =A(x,t)\exp[i(kx-\omega t)]+c.c., \Lambda_{2} & =B(x,t)\exp[i(kx-\omega t)]+c.c.,
\end{eqnarray}

and plugging these relations in the equations together, with account of %
\eqref{approximationsA}, linear independence of complex conjugated parts and
strong inequality \eqref{approximationsA}, leaving the nonlinear resonant
terms in the r.h.s, one can obtain
\begin{equation}
c(ikA_{t}-i\omega A_{x})=-\frac{\chi _{e}^{(3)}c^{2}\omega ^{6}}{16p^{4}q^{2}}%
[A(3|A|^{2}+2|B|^{2})+A^{\ast }B^{2}],  \label{eqA-full}
\end{equation}%
\begin{equation}
c(ikB_{t}-i\omega B_{x})=-\frac{\chi _{e}^{(3)}c^{2}\omega ^{6}}{16p^{4}q^{2}}%
[B(2|A|^{2}+3|B|^{2})+A^{2}B^{\ast }].  \label{eqB-full}
\end{equation}%
The parameter of the solution $k$ is chosen to simplify the equations as
\begin{equation}
-ci\omega (ik)A+pqA=(k\omega +\frac{pq}{c})A=0,
\end{equation}%
that is equivalent to the expression
\begin{equation}
k=-\frac{pq}{c\omega },
\end{equation}%
that fix the phase velocity of the carrier wave as in linear case. Equations %
\eqref{eqA-full} and \eqref{eqB-full} with account approximation %
\eqref{approximationsA} are:
\begin{eqnarray}
\frac{k}{\omega }A_{t}-A_{x} &=&-i\frac{\chi _{e}^{(3)}c\omega ^{5}}{%
16p^{4}q^{2}}[A(3|A|^{2}+2|B|^{2})+B^{2}A^{\ast }],  \label{sysinter} \\
\frac{k}{\omega }B_{t}-B_{x} &=&-i\frac{\chi _{e}^{(3)}c\omega ^{5}}{%
16p^{4}q^{2}}[B(2|A|^{2}+3|B|^{2})+A^{2}B^{\ast }].
\end{eqnarray}%
This system of equations describes the interaction between orthogonal
polarization modes, propagating to the left in metamaterials.

\subsection{Particular solution of \eqref{sysinter}}
After change the variables:
\begin{equation}
\eta =x-\frac{t \omega}{k},\xi =x+\frac{t\omega}{k},
\end{equation}%
\begin{equation*}
\partial _{x}=\partial _{\xi }+\partial _{\eta },\partial_{t}=\frac{\omega}{k} \partial _{\xi }-\frac{\omega}{k}\partial _{\eta },A(x,t)\rightarrow
A(\eta ,\xi ), B(x,t) \rightarrow B(\eta ,\xi ),
\end{equation*}
\begin{eqnarray}
2 A_{\eta} &=&i\frac{\chi _{e}^{(3)}c\omega ^{5}}{16p^{4}q^{2}}[A(2|A|^{2}+3|B|^{2})+B^{2}A^{\ast }], \label{sysABB} \\
2 B_{\eta} &=&i\frac{\chi _{e}^{(3)}c\omega^{5}}{16p^{4}q^{2}}[B(3|A|^{2}+2|B|^{2})+A^{2}B^{\ast }].
\end{eqnarray}
\begin{eqnarray}
  A = A_0 \exp(i\varphi_1 (\eta)),   B = B_0 \exp(i\varphi_2 (\eta)),
\end{eqnarray}
\begin{eqnarray}
 2\varphi_1'   &=&\frac{\chi _{e}^{(3)}c\omega ^{5}}{%
16p^{4}q^{2}}[ (3|A_0|^{2}+2|B_0|^{2})  +B_0^{2}\frac{A_0^{\ast}}{A_0} \exp(i(2\varphi_2(\eta) -2\varphi_1(\eta) ))  ]  ,  \\
 2\varphi_2'   &=&\frac{\chi _{e}^{(3)}c\omega ^{5}}{%
16p^{4}q^{2}}[ (2|A_0|^{2}+3|B_0|^{2})  +A_0^{2}\frac{B_0^{\ast }}{B_0} \exp(i(2\varphi_1(\eta) - 2 \varphi_2(\eta)))].
\end{eqnarray}
Substraction the second equation from the first one and mark
$$2\varphi_1 - 2\varphi_2 \equiv z(\eta),$$
\begin{eqnarray}
z' &=&\frac{\chi _{e}^{(3)}c\omega ^{5}}{%
16p^{4}q^{2}}[ |A_0|^{2}-|B_0|^{2}  +B_0^{2}\frac{A_0^{\ast}}{A_0} \exp(iz(\eta)) +A_0^{2}\frac{B_0^{\ast }}{B_0} \exp(-iz(\eta))].
\end{eqnarray}

\subsubsection{case $\varphi_1 (\eta)  = \varphi_2 + \text{C}$ }
where:
$$C = \text{const}, \exp(2iC) \equiv y $$
C means the phase shift.
There are two solutions \eqref{sysABB} in this case. \textcolor{red}{We consider only the one with phase shift:}
\begin{eqnarray}
  A_2 = A_0 \exp\left\{i\frac{\chi _{e}^{(3)}c\omega ^{5}}{%
32p^{4}q^{2}}(|A_0|^{2}+4|B_0|^{2})\eta - \frac{i}{2}\ln \left[ \frac{B_0}{ A_0^{2} B_0^{\ast}} (|A_0|^{2} -|B_0|^{2} ) \right]\right\}, \\
  B_2 = B_0 \exp(i\frac{3 \chi _{e}^{(3)}c\omega ^{5}}{32p^{4}q^{2}}(|A_0|^{2}+|B_0|^{2})\eta ).
\end{eqnarray}
\begin{figure}[h]
\begin{center}
\begin{minipage}[H]{0.4\linewidth}
\includegraphics[scale=0.49]{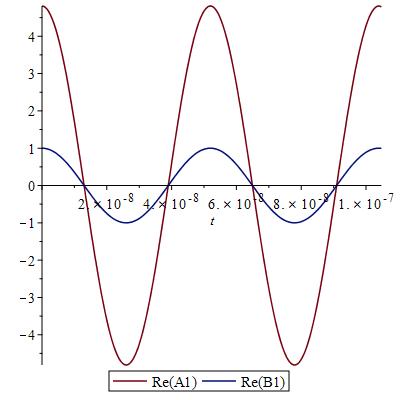}
\end{minipage}
\end{center}
\caption{Real parts of functions $A_2(x,t)$, $B_2 (x,t)$ for $p=10^9 Hz, q = 10^3p , \omega = 1.2q, \chi=10^{-20} \, \frac{m^2}{V^2}, A_0 = B_0 = 1 \, T, x = 2 \, m$ }
\end{figure}
The phase shift is:
$$C_2 =- \frac{1}{2}\ln \left[ \frac{B_0}{ A_0^{2} B_0^{\ast}} (|A_0|^{2} -|B_0|^{2} ) \right] = - \frac{1}{2}\ln \left[ \frac{B_0 A_0^{\ast}}{ A_0 B_0^{\ast}}  - \frac{B_0^2}{ A_0^{2}} \right] .$$
To plot the dependence of the shift on amplitude ratio, we mark $\frac{B_0}{A_0}$ as $z$. Then:
$$C_2 =- \frac{1}{2}\ln \left[ \frac{B_0 A_0^{\ast}}{ A_0 B_0^{\ast}}  - \frac{B_0^2}{ A_0^{2}} \right]  \rightarrow - \frac{1}{2} \ln ( \frac{z}{z^{\ast}} - z^2 ).$$
\begin{figure}[h]
\begin{center}
\begin{minipage}[H]{0.4\linewidth}
\includegraphics[scale=0.4]{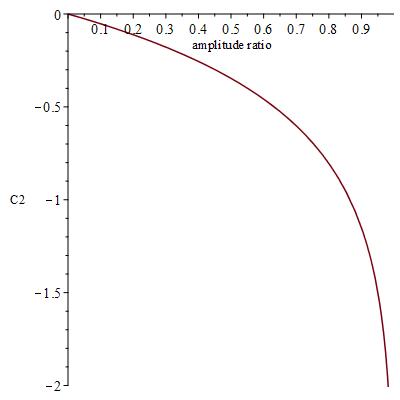}
\end{minipage}
\end{center}
\caption{Phase shift $C_2$ in dependence of amplitude ratio $\frac{B_0}{A_0}$ for real or imaginary  $A_0$ and $B_0$}
\end{figure}
\section{Traveling wave solution of \eqref{systemdifft} }

We return to \eqref{systemdifft}:
\begin{eqnarray}
\frac{c}{pq}\partial _{xt}\Lambda _{1}+\Lambda _{1} &=&\frac{1}{16}\chi
_{e}^{(3)}\frac{c^2}{p^4q^2}[(\partial _{t}^2\Lambda
_{1})^3+\partial _{t}^2 \Lambda _{1}(\partial _{t}^2 \Lambda_2 )^2],
\\
\frac{c}{pq}\partial _{xt}\Lambda_2 +\Lambda_2  &=&\frac{1}{16}\chi
_{e}^{(3)}\frac{c^2}{p^4q^2}[(\partial _{t}^2 \Lambda_2
)^3+\partial _{t}^2 \Lambda_2 (\partial _{t}^2\Lambda _{1})^2].
\end{eqnarray}
 After transition to variables:
\begin{equation}
x = \eta, \xi = x + v t, \label{xxi}
\end{equation}
where $v$ - parameter with dimension of velocity,and apply the approximation.
$$\partial_{\eta} L_1 = \partial_{\eta} L_2 = 0,$$
and then dividing all on $ \frac{vc}{pq}$ :
\begin{eqnarray}
\partial_\xi^2  L_1 + \kappa  L_1 =  \lambda_e^2  [(\partial_\xi^2 L_{1})^3+\partial_\xi^2 L_{1}(\partial_\xi^2 L_2)^2] ,  \\
\partial_\xi^2  L_2 + \kappa  L_2 = \lambda_e^2  [(\partial_\xi^2 L_2
)^3+\partial_\xi^2 L_2(\partial_\xi^2 L_{1})^2].
\end{eqnarray}
$\kappa, \lambda_e^2$   - parameters :
\begin{eqnarray}
\kappa = \frac{pq}{vc}, \label{kappa} \\
\lambda_e^2 =  \frac{v^5}{16}\chi^{(3)}\frac{c}{p^3q}. \label{lamnl}
\end{eqnarray}
Reduction
\begin{equation}\label{red}
  L_1 =  L_2 = L,
\end{equation}
and division on $2 \lambda_e^2$ yield:
\begin{equation}
  (\partial_\xi^2 L)^3 - \frac{1}{2 \lambda_e^2}\partial_\xi^2  L -  \frac{\kappa}{2 \lambda_e^2}  L = 0.
\end{equation}
Solving this cubic equation by Cardano formula with respect to $\partial_\xi^2  L$, accounting only real solution and expanding it in power series on $L$, we obtain the approximate equation:
\begin{equation}
  (\partial_\xi^2  L)_1 + 2 \kappa^3 \lambda_e^2 L^3 + \kappa L = 0 ,
\end{equation}
After rescaling of variables and accounting relations (\ref{kappa},\ref{lamnl}):
\begin{equation}
  \partial_\xi^2  L + \frac{1}{8} \frac{q^2 v^2 \chi^{(3)}}{c^2} L^3 + \frac{pq}{vc} L = 0 ,
\end{equation}
Introducing amplitude parameter $\alpha$, dimensionless function $l$ as:
$$L = \alpha l,$$
and dimensionless variable $\beta$ as:
$$ \xi = n \beta,$$
we obtain an equation on $l$:
\begin{equation}
  l_{\beta \beta} + \frac{pq}{vc} n^2 l + \frac{1}{8} \frac{q^2 v^2 \chi^{(3)}}{c^2} n^2  \alpha^2 l^3  = 0 ,
\end{equation}
Comparison with equation for elliptic cosine cn$(u,m)$:
$$y'' + (1 - 2m^2) y + 2 m^2 y^3 = 0, $$
leads to the system of equation relative $m$ and $v$:
\begin{eqnarray}
  1 - 2m^2 &=& \frac{pq}{vc} n^2,  \\
   \frac{1}{8} \frac{q^2 v^2 \chi^{(3)}}{c^2} \alpha^2 n^2   &=& 2 m^2.
\end{eqnarray}
From this system it's easy to find $m$:
\begin{equation}\label{sol:m}
  m^2 = \frac{1}{16} \frac{q^2 v^2 \chi^{(3)}}{c^2} \alpha^2 n^2 ,
\end{equation}
After doing some algebra, one can easily derive the cubic equation on $v$
\begin{equation}\label{eqforv}
v - \frac{1}{8} \frac{q^2  \chi^{(3)}}{c^2} \alpha^2 n^2 v^3 - \frac{pq}{c} n^2 = 0.
\end{equation}
Solution by Cardano formula can lead to wrong results for case $\alpha \rightarrow 0$. Hence, for small amplitudes we need to look for solution using the  successive approximations method.  In first approximation $\alpha \rightarrow 0$:
\begin{equation}\label{v0}
v_0 = \frac{pq}{c} n^2
\end{equation}
Substitution it into \eqref{eqforv} yields:
\begin{equation}\label{vapprox}
v_1 \approx   \frac{1}{8} \frac{q^2  \chi^{(3)}}{c^2} \alpha^2 n^2 \frac{p^3q^3}{c^3} n^6 + \frac{pq}{c} n^2.
\end{equation}
In small amplitude approximation $\Lambda_2$-wave is:
\begin{equation}\label{L2fin}
\Lambda_2 = \alpha \text{cn} (\frac{x+v_1 t}{n}, \frac{1}{4} \frac{q v \sqrt{\chi^{(3)}}}{c} \alpha n  ),
\end{equation}
where $v$ and $\alpha$ are velocity and amplitude of induced wave.  $\frac{v}{n}$ is a frequency of nonlinear wave on boundary ($x=0$).

Changing $\alpha, n  $ we can change the behaviour of the wave on boundary.

\section{Conclusion}

In this work the wave propagation of two polarizations in 1D-metamaterial
has been studied. The general equation of directed wave propagation in
1D-metamaterial with two polarization has been obtained. It's shown for
Drude metamaterial with Kerr nonlinearity it has the form of "vector SPE".

The system of equations \eqref{sysinter} which describe the interaction between orthogonal polarization modes, propagating to the left in metamaterials for case of slow-varying envelopes has been obtained.

\small{

}

\begin{thebibliography}{99}
\bibitem{Zakh} {\small V. I. Talanov, ZhETF Pis. Red. 2, 223 (1965)[JETP
Lett. 2, 141 (1965)]. Zakharov V.E. J. Appl. Mech. Tech. Phys. 9, 190
(1968). }

\bibitem{ZS} {\small Zakharov, V.E. and Shabat, A.B. (1971) Exact theory of
two-dimensional self- focusing and one-dimensional modulation of waves in
nonlinear media, Zhurn. Eksp. Teor. Fiz. 61, 118-134 [(1972) Sov. Phys. JETP
34, 62-69]. }

\bibitem{DL} {\small E. Doktorov S.B. Leble Dressing method in mathematical
physics. ( Springer-Verlag, 2007) }

\bibitem{SU} {\small Sazonov S. V., Ustinov N. V. New class of extremely
short electromagnetic solitons, Pis'ma v Zh. Eksper. Teoret. Fiz., 83:11
(2006), 573-578 General class of the traveling waves propagating in a
nonlinear oppositely-directional coupler }

\bibitem{SW} {\small Sch\"afer T., Wayne C.E. Propagation of ultra-short
optical pulses in cubic nonlinear media.Phys. D 196, 90-105 (2004) }

\bibitem{SW2} {\small Chung Y., Jones C.K.R.T., Sch\"afer T., Wayne C.E.
Ultra-short pulses in linear and nonlinear media. // Nonlinearity, 18. -
2005, P. 1351-1374 }

\bibitem{K} {\small P. Kinsler, Phys. Rev. A \textbf{81}, (2010), 023808. }

\bibitem{L} {\small S. Leble. Nonlinear Waves in Waveguides (Springer,
Heidelberg, 1990). }

\bibitem{Dobr} {\small V.V.Belov, S.YU.Dobrokhotov, T.YA.Tudorovskiy.
Operator separation of variables for adiabatic problem in quantum and wave
mechanic // Journal of Engineering Mathematics (2006) }

\bibitem{Pi} {\small S. Pitois, G. Millot, S. Wabnitz ,Nonlinear
polarization dynamics of counterpropagating waves in an isotropic optical
fiber: theory and experiments, J. Opt. Soc. Am. B/ Vol. 18, No. 4/ April
2001 }

\bibitem{KL} {\small M. Kuszner, S. Leble, Directed Electromagnetic Pulse
Dynamics: Projecting Operators Method J. Phys. Soc. Jpn. 80 (2011) 024002. }

\bibitem{P} {\small A. A. Perelomova , Projectors in nonlinear evolution
problem: acoustic solitons of bubbly liquid, Applied Mathematics Letters, 13
(2000), 93-98; Nonlinear dynamics of vertically propagating acoustic waves
in a stratified atmosphere , Acta Acustica, 84(6) (1998), 1002-1006. }

\bibitem{Pe} {\small A. Perelomova, \textit{Development of linear projecting
in studies of non-linear flow. Acoustic heating induced by non-periodic
sound }, Phys. Lett. A 357,2006, 42-47. }

\bibitem{KuLe} {\small M.Kuszner, S.Leble, Ultrashort Opposite Directed
Pulses Dynamics with Kerr Effect and Polarization Account Journal of the
Physical Society of Japan 83 (2014) }

\bibitem{PKB} {\small M. Pietrzyk, I. Kanattsikov, and U. Bandelow, "On the
propagation of vector ultra-short pulses", J. of Nonlin. Math. Phys. 15, 2,
2008 }

\bibitem{Ziol} {\small R.W. Ziolkowski and A. Kipple. Causality and
double-negative metamaterials Phys. Rev. E, vol.68, 026615, Aug. 2003 }

\bibitem{Ziol2} {\small R.W. Ziolkowski and F. Auzanneau. Passive artificial
molecule realizations of dielectric materials, J. Appl. Phys., vol.82,
pp.3195-3198, Oct. 1997 }

\bibitem{Kan2} {\small M. Pietrzyk, I. Kanattsikov, On the Generalized Short
Pulse Equation Describing Propagation of Few-Cycle Pulses in Metamaterial
Optical Fibers // Theoretical Physics and Its Applications, Moscow, 2013. }

\bibitem{AmLe} {\small D. Ampilogov, S. Leble, General Equation for Directed
Electromagnetic Wave Propagation in 1D Metamaterial: Projecting Operator
Method. TASK Quarterly, V.20, No.2 (2016) }

\bibitem{Kaplan} {\small A. E. Kaplan, Light-induced nonreciprocity, field
invariants, and nonlinear eigenpolarizations, Opt. Lett. 8, pp.560- 562
(1983) }

\bibitem{Christos2} {\small Christos Argyropoulos, et. al. Enhanced
Nonlinear Effects in Metamaterials and Plasmonics // Advanced
Electromagnetics, Vol. 1, No. 1, May 2012, pp. 46-51 }

\bibitem{Zhaqilao} {\small Z.Zhaqilao, Q.Hu, Z. Qiao, Multi-soliton
solutions and the Cauchy problem for a two-component short pulse system //
Nonlinearity, V.30, No.10, 2017}

\bibitem{AmLETASK} {\small D.Ampilogov. Interaction of orthogonal-polarized waves in 1D-metamaterial // TASK QUARTERLY vol. 21, No 2, 2017, pp. 605-619}

\end{thebibliography}
\end{document}